\documentclass{aa}  

\usepackage{float}
\usepackage{graphicx}
\usepackage{txfonts}
\usepackage{natbib}
\bibpunct{(}{)}{,}{a}{}{,}  
\usepackage{pdflscape}
\usepackage{hyperref}
\hypersetup{breaklinks=true,
            colorlinks,
            filecolor=blue,
            linkcolor=blue,
            citecolor=blue,
            urlcolor=blue,
            anchorcolor=blue} 
\usepackage{listings}
\begin{document}

   \title{\texttt{pastamarkers}: astrophysical data visualization with pasta-like markers}


   \author{PASTA Collaboration
          \and
          N. Borghi
          \inst{1,2}
          \and   
          E. Ceccarelli
          \inst{1,2}            
          \and
          A. Della Croce
          \inst{1,2}
          \and            
          L. Leuzzi
          \inst{1,2}
          \and     
          L. Rosignoli
          \inst{1,2}   
          \and           
          A. Traina
          \inst{1,2}            
          }

   \institute{Department of Physics and Astronomy, University of 
             Bologna, Via Gobetti 93/2, 40129 Bologna, Italy
         \and
             INAF - Astrophysics and Space Science Observatory of 
             Bologna, Via Gobetti 93/3, 40129 Bologna, Italy 
             }

   \date{Received xxxx; accepted yyyy}

 
  \abstract
   {}
   {We aim at facilitating the visualization of astrophysical data for several tasks, such as uncovering patterns, presenting results to the community, and facilitating the understanding of complex physical relationships to the public.}
   {We present \texttt{pastamarkers}, a customized \texttt{Python} package fully compatible with \texttt{matplotlib}, that contains unique pasta-shaped markers meant to enhance the visualization of astrophysical data.}
   {We prove that using different pasta types as markers can improve the clarity of astrophysical plots by reproducing some of the most famous plots in the literature.}
   {}


   \maketitle
%

\section{Introduction}\label{sec:intro}
The analysis of astrophysical data presents a formidable challenge due to the intrinsic complexities of the underlying physical phenomena being studied. Astrophysical datasets typically entail vast volumes of information marked by numerous degeneracies and intricate patterns. These complexities can arise from various factors such as numerical computation, observational limitations, instrumental noise, and/or the intrinsic variability of astrophysical sources themselves. Consequently, unraveling true relationships within these datasets requires sophisticated data analysis tools and visualization techniques.

Data visualization in particular emerges as an indispensable tool in navigating the complexity of astrophysical signals. While statistics is the pillar of the scientific method, visual representation offers a means to recognize general properties and get an intuition of trends between different quantities. This step of the job is often underestimated, but it will get more and more important as the volume of data available in the field increases. Given the recent and upcoming launch of imaging and spectroscopic surveys, the volume of observational data incoming is bound to explode: getting a first glimpse of the general properties of the data will be of great help as a first step before proceeding to more accurate analysis. Moreover, visualization aids in the identification of outliers, clusters, and correlations, thereby facilitating deeper insights into the astrophysical processes. Data visualization is a bridge between raw data and meaningful interpretations, enabling scientists to explore and understand complex data more effectively. Finally, presenting the results to the community and the general public, which is also an important part of research, can benefit from clear visualization of plots.

Despite the critical role of data visualization in astrophysics, the field currently lacks a tailored set of markers, that would enhance the unique characteristics of astrophysical data. Given the intricate nature of the observations and the multitude of parameters involved, existing visualization tools often fall short of capturing the nuances of the datasets. Thus, there is a pressing need to develop specialized markers and visualization techniques optimized for astrophysical research. Such tools would improve the clarity and interpretability of astrophysical data, and also facilitate the discovery of new physics in our Universe.

In this work, we explore using unconventional markers when showcasing astrophysical plots as a way of better conveying the information in the data and potentially making the message clearer for the general public. To do so, we develop a comprehensive new set of markers featuring shapes inspired by various types of pasta. This choice stems from the compelling desire to blend two beloved aspects of human culture: science and Italian cuisine. We make the Promoting Astrophysical Studies Through Aliments (PASTA) markers available to the community.

This paper is organized as follows: in Section \ref{sec:pastamarker} we present the contents of the package and explain how to use it, in Section \ref{sec:examples} we show a few useful examples of the PASTA markers, by reproducing iconic plots found in the literature, and we conclude in Section \ref{sec:conclusions}.

\section{\texttt{pastamarkers}}\label{sec:pastamarker}

\begin{figure}[!th]
    \centering
    \includegraphics[width=.49\textwidth]{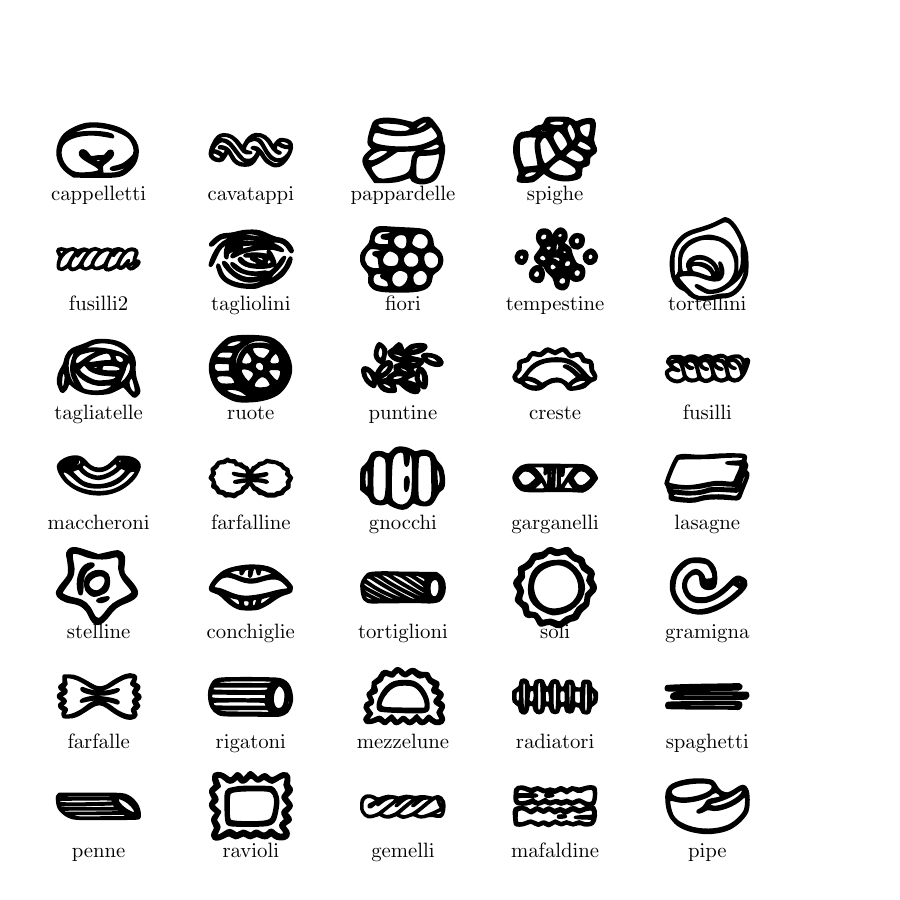}
    \caption{All PASTA markers included within the \texttt{pastamarkers} package.}
    \label{fig:pastas_yummy}
\end{figure}

\subsection{Brief description}
The \texttt{pastamarkers} package is an open source code available at \url{https://github.com/LR-inaf/pasta-marker}.

This package exploits the \href{https://pypi.org/project/svgpath2mpl/}{svgpath2mpl} library to convert svg files into \href{https://matplotlib.org/stable/api/path_api.html}{matplotlib Path object}, storing the Path collection into \texttt{Python} files callable into the main \texttt{Python} script, where the plots are created.

In Figure \ref{fig:pastas_yummy}, we showcase all 34 new markers included in the \texttt{pastamarkers} package, each accompanied by its corresponding name listed below. In future implementations of the current package, we envision additional improvements. Specifically, we plan to pair each PASTA marker with its ideal sauce accompaniment (i.e. marker color), along with instructions for configuring the plot with utensils and glassware.

\subsection{Package usage}
In order to use the package, you need to download the repo at this \href{https://github.com/LR-inaf/pasta-marker}{link}. Then, two options can be pursued:
\begin{itemize}
    \item Add to your \texttt{Python} script the path to the \emph{pastamarkers} folder within the repo like:
    \begin{lstlisting}[language=python]
    import sys
    sys.path.append(path_to_pastamarker)
    \end{lstlisting}

    \item Copy the \emph{pastamarkers} folder into the \emph{site-packages} folder of your \texttt{Python} environment.
\end{itemize}
\section{Examples}\label{sec:examples}
\begin{figure}[!th]
    \centering
    \includegraphics[width=.49\textwidth]{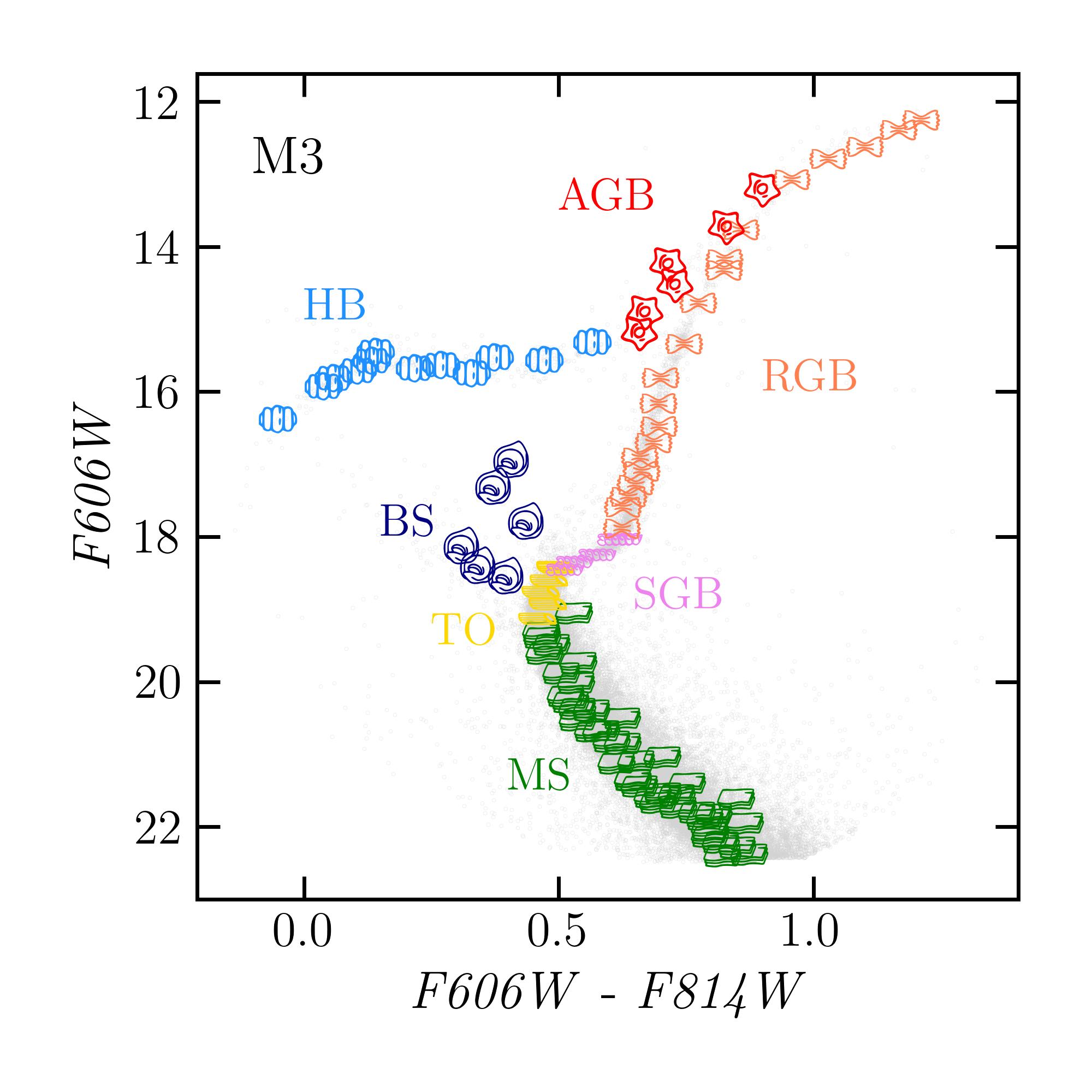}
    \caption{CMD of the globular cluster M3. Different evolutionary stages are highlighted with different colors and with the different PASTA markers introduced in this work.}
    \label{fig:cmd}
\end{figure}
\begin{figure}[!th]
    \centering
    \includegraphics[width=.49\textwidth]{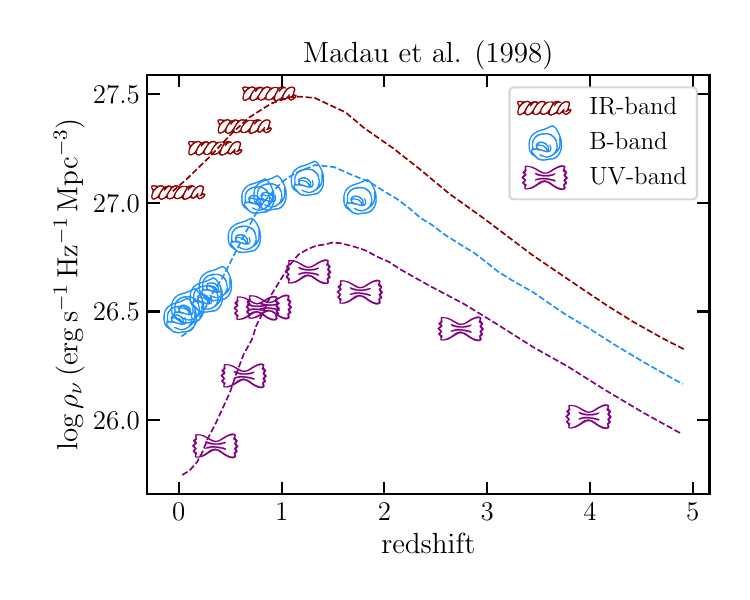}
    \caption{
    Redshift evolution of the luminosity density of the Universe. Data in the IR, B, and UV restframe bands are from \citet{madau_pozzetti_dickinson1998} shown using different PASTA markers.
    }
    \label{fig:mpd_plot}
\end{figure}

In this Section, we show the application of the new set of markers in generating visually appealing plots while maintaining scientific integrity. Specifically, we demonstrate how traditional plots, though historically significant for the astrophysical community, can be rendered even more iconic and comprehensible through the use of unconventional markers.

In Figure \ref{fig:cmd}, we present the color-magnitude diagram (CMD) of the globular cluster M3 (NGC 5272) in a completely new fashion. To achieve this magnificent result, we revised the famous version displayed in \citet{renzini1988} by employing high-quality photometric data acquired under the Hubble Space Telescope (HST) Legacy Survey of Galactic Globular Clusters \citep[HUGS,][]{piotto2015}. The gray background points represent all cataloged M3 stars available in the dataset published by \citet{piotto2015}. To better visualize the numerous stellar evolutionary stages, we selected only stars with membership probability $> 90\%$ and rms $< 0.005$ in both \textit{F606W} and \textit{F814W} bands. 

Figure \ref{fig:mpd_plot} is an updated version of the historic plot by \citet{madau_pozzetti_dickinson1998}. Here, we shows model predictions for the redshift evolution of the luminosity density $\rho_{\nu}$ of the Universe. Different PASTA markers are used to plot data taken in IR- (brown fusilli2), B- (blue tortellini) and UV- (purple farfalle) restframe bands.

\section{Conclusions}\label{sec:conclusions}
In this work, we have presented the \texttt{pastamarkers} package, an innovative \texttt{matplotlib}-based library that provides a wide range of pasta varieties that can be used as markers when doing astrophysical plots. Through meticulous experimentation and analysis, we have curated a focused selection of pasta types, each uniquely suited to enhance the visual representation of astrophysical data. Our investigation involved revisiting seminal plots from the astrophysical literature, where we seamlessly integrated our newly introduced PASTA markers. The results were compelling, demonstrating the remarkable efficacy of our markers in effectively conveying complex astrophysical information with clarity and precision. We are pleased to announce that the source code for \texttt{pastamarkers} is now publicly available through the provided \href{https://github.com/LR-inaf/pasta-marker}{link}. Alongside the code, we have included comprehensive documentation and examples to facilitate the integration and usage of the package in astrophysical research and visualization endeavors.

\begin{acknowledgements}

EC and ADC are grateful to D.Massari and L.Scaloni for their helpful comments to reproduce Figure \ref{fig:cmd}.

\end{acknowledgements}

\bibliographystyle{aa}
\bibliography{aanda.bib}

\end{document}